\documentclass[journal=nalefd,manuscript=letter,layout=twocolumn]{achemso}

\usepackage[version=3]{mhchem} 

\pdfoutput=1

\usepackage[english]{babel}
\usepackage[utf8]{inputenc}
\usepackage{newtxtext, newtxmath}
\usepackage{soul}
\usepackage[scaled=0.8]{beramono}
\usepackage[T1]{fontenc}
\usepackage{breqn}

\usepackage{acronym}
\usepackage{bm}
\usepackage{dsfont}
\usepackage{graphicx}
\usepackage{mathtools}
\usepackage{microtype}
\usepackage{tikz}

\usetikzlibrary{decorations.markings}
\usetikzlibrary{decorations.pathmorphing}

\definecolor{mauve}{RGB}{94, 60, 153}

\usepackage[colorlinks, allcolors=mauve]{hyperref}
\usepackage[figure, figure*]{hypcap}

\newcommand*{\IPM}[0]{{School of Nano Science, Institute for Research in Fundamental Sciences (IPM), 19395-5531 Tehran, Iran}}

\newcommand*{\DIPC}[0]{{Donostia International Physics Center (DIPC), 20018 Donostia-San Sebasti\'an, Spain}}

\newcommand*{\IFS}[0]{{Centre for Advanced Laser Techniques, Institute of Physics, 10000 Zagreb, Croatia}}

\author{Zahra Torbatian}
\affiliation{\IPM}
\author{Dino Novko}
\email{dino.novko@gmail.com}
\affiliation{\IFS}
\altaffiliation{\DIPC}

\title{Plasmon Excitations Across the Charge-Density-Wave Transition in Single-Layer TiSe$_2$}


\begin{document}

\sloppy

\begin{abstract}
$1T$-TiSe$_2$ is believed to posses a soft electronic mode, i.e., plasmon or exciton, that might be responsible for the exciton condensation and charge-density-wave (CDW) transition. Here, we explore collective electronic excitations in single-layer $1T$-TiSe$_2$ by using the \emph{ab-initio} electromagnetic linear response and unveil intricate scattering pathways of two-dimensional (2D) plasmon mode near the CDW phase. We found the dominant role of plasmon-phonon scattering, which in combination with the CDW gap excitations leads to the anomalous temperature dependence of the plasmon linewidth across the CDW transition. Below the transition temperature $T_{\rm CDW}$ a strong hybridization between 2D plasmon and CDW excitations is obtained. These optical features are highly tunable due to temperature-dependent CDW-related modifications of electronic structure and electron-phonon coupling and make CDW-bearing systems potentially interesting for applications in optoelectronics and low-loss plasmonics.
\end{abstract}

\vspace{10mm}

Transition metal dichalcogenides (TMDs) host a variety of correlated ordered states which makes them an ideal playground for exploring and manipulating different fundamental interactions in condensed matter\,\cite{rossnagel11,manzeli17}. Quasi-two-dimensional $1T$-TiSe$_2$ belongs to this category having a rich phase diagram, including unconventional charged density wave (CDW)\,\cite{disalvo76} and superconductivity\,\cite{morosan2006superconductivity}, tunable with temperature\,\cite{disalvo76}, pressure\,\cite{kusmartseva2009pressure}, and doping\,\cite{morosan2006superconductivity,luo16}. The origin of the CDW order in TiSe$_2$ is still actively debated, where the proposed underlying mechanisms range from purely electronic\,\cite{disalvo76,cercellier07,kogar17}, purely phononic\,\cite{zhou20}, and the combination of both\, \cite{yoshida80,holt01,calandra11,hedayat19,lian20,novko22}. In the former case, the ordered state is stabilized by the soft electronic mode\,\cite{kogar17}, i.e., exciton or plasmon, which constitutes the intriguing excitonic insulator scenario\,\cite{jerome67}. However, despite being observed in TiSe$_2$, the role of the soft plasmon in the CDW formation is unclear, especially since it is strongly Landau damped in the relevant momentum region\,\cite{lin22}.

The interplay of plasmons and correlated states can result in some unexpected and desirable optoelectronic properties\,\cite{vanloon14}, like tunable, long-lived and flat correlated plasmon modes\,\cite{asmara17,gao22}. In bulk $2H$ TMDs, such as TaSe$_2$, TaS$_2$, NbSe$_2$, the experimentally observed negative plasmon dispersion was explained in terms of coupling with the CDW state\,\cite{wezel11,konig13}. Furthermore, plasmon mode in bulk TiSe$_2$ was shown to be highly modified across the CDW transition due to the CDW gap excitations\,\cite{li2007semimetal,kogar17,lian2019charge,lin22}. The two-dimensional (2D) plasmon modes in atomically thin TMDs\,\cite{andersen13,gjerding17,jornada20} are characterized with long-wavelength gapless dispersion, low losses, and broad spectral range, opening further possibilities for the CDW-plasmon coupling. For instance, optical measurements of TaSe$_2$ thin films below transition temperature $T_{\rm CDW}$ found intricate hybrid mode consisting of 2D plasmon and CDW excitations showing anomalous broadening\,\cite{song2021plasmons}. It is therefore clear that, besides the doping\,\cite{fei12,novko17} and dielectric environment\,\cite{despoja19}, controlling the CDW order is highly appealing pathway for tuning the plasmonic properties, and that the CDW-bearing 2D materials are potentially desirable for applications in plasmonics and optoelectronics. Hence, the corresponding microscopic description of coupling between plasmon and CDW are not only essential for unveiling the origin of the CDW formation, but also for busting the optical properties in TMDs.

Here, we investigate the influence of CDW transition on the electron excitation spectra of the TiSe$_2$ monolayer by means of the density functional perturbation theory (DFPT)\,\cite{baroni01} and current-current linear response formalism\,\cite{Novko2016,novko17}. The present approach is able to disentangle the relevant plasmon damping channels, such as electron-phonon scattering and Landau damping due to CDW gap excitations. The results show how the CDW-related structural distortions impact the electronic bands and phonon energies, which in turn leads to remarkable modifications of 2D plasmon. 
We show that the coupling of plasmon with phonons (in particular, with soft CDW phonon) is responsible for the drastic increase of the plasmon decay rate for $T>T_{\rm CDW}$, while the closing of the CDW electronic gap, and accompanying increase of the Drude weight, for the plasmon energy increase.
In fact, in accordance with the optical conductivity measurements\,\cite{li2007semimetal}, we show that the CDW interband excitations are suppressed above $T_{\rm CDW}$, which in combination with the plasmon-phonon scattering channel leads to the anomalous temperature-dependence of the plasmon linewidth\,\cite{li2007semimetal,lin22}. Interestingly, below $T_{\rm CDW}$ these CDW exictations are more pronounced and are coupled to the 2D plasmon forming a hybrid CDW-plasmon mode, in a close resemblance to the coupled mode recently found in TaSe$_2$\,\cite{song2021plasmons}.

The ground state electronic structure calculations are performed by means of the plane-wave density-functional-theory (DFT) code \textsc{Quantum Espresso}\,\cite{qe09} and by using the semi-local exchange-correlation PBE functional. Phonon dynamics and electron-phonon coupling (EPC), which are found to be crucial for the CDW properties in TiSe$_2$\,\cite{calandra11,hellgren2017critical}, are obtained within the DFPT\,\cite{baroni01} framework, and then interpolated with maximally-localized Wannier functions\,\cite{wan90} and \textsc{EPW} code\,\cite{ponce2016epw}. Plasmon polariton dispersion and optical conductivity are calculated with \emph{ab-initio} current-current linear response method suitable for simulating transverse and longitudinal optical response of 2D materials\,\cite{Novko2016}, which additionally includes higher-order electron-phonon scattering contributions\,\cite{novko17}. This allows for the accurate assessment of the plasmon damping and energy renormalization due to EPC\,\cite{novko17,torbatian2020tunable,torbatian2021hyperbolic}. Further computational details can be found in the Supporting Information (SI).

\begin{figure}[!t]
\begin{center}
\includegraphics[width=\columnwidth]{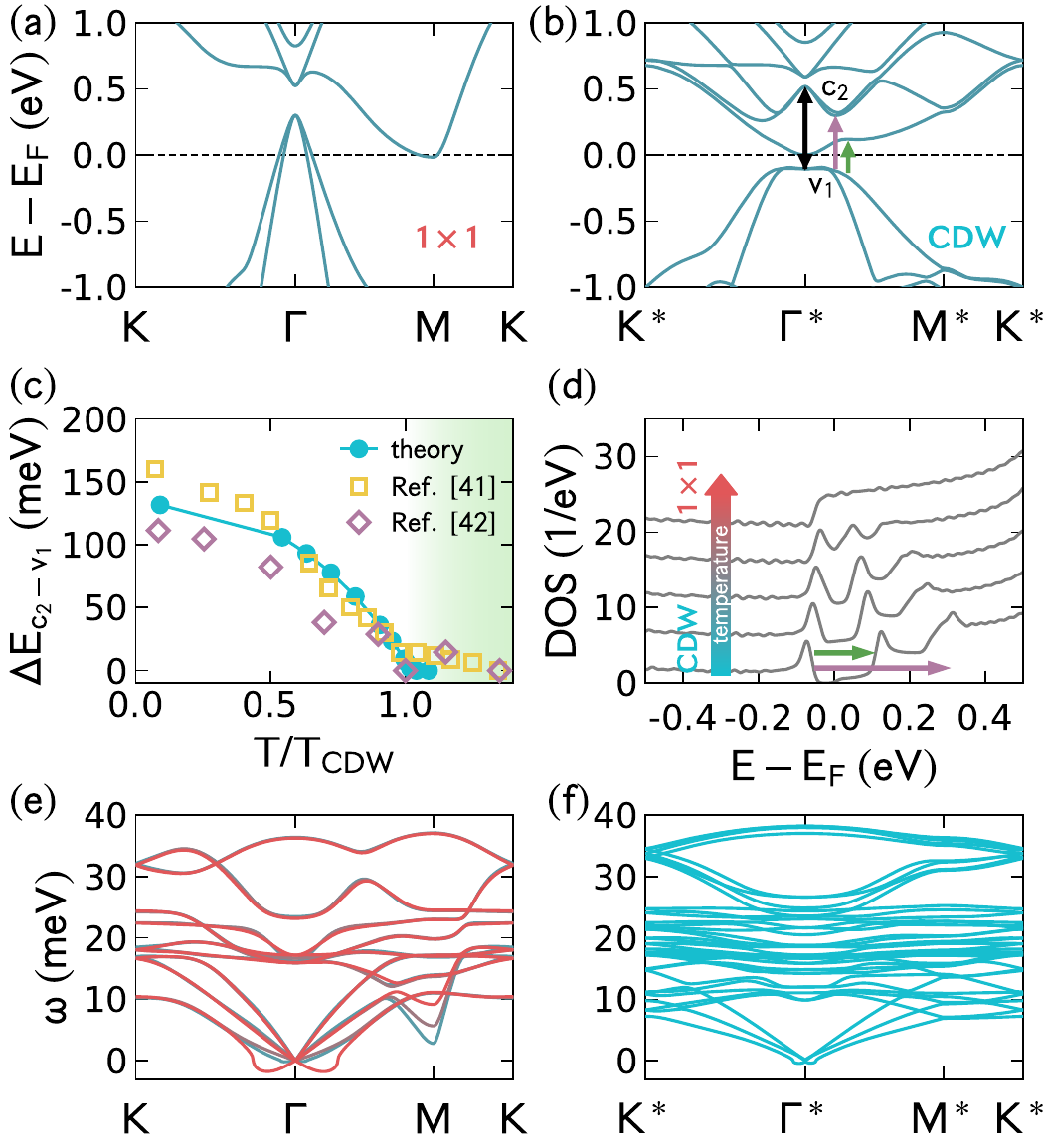}
\caption{Electronic band structure of $1T$-TiSe$_2$ along high-symmetry points of the Brillouin zone for (a) $1\times 1$ phase at high-temperatures $T>T_{\rm CDW}$ and (b) $2\times 2$ CDW structure with periodic lattice distortions at $T\ll T_{\rm CDW}$. The asterisk sign for high-symmetry points denotes the reconstructed $2\times 2$ Brillouin zone. The black arrow denotes the CDW gap opening at $\mathrm{k}=\Gamma^{\ast}$, while purple and green arrows show electronic transitions with highest density of states. (c) Relative temperature shifts of the CDW gap at $\mathrm{k}=\Gamma^{\ast}$, $\Delta E_{c_2-v_1}$, as obtained with DFT and experiments\,\cite{monney10,monney12} with respect to $E_{c_2-v_1}(T\gg T_{\rm CDW})$. Note that the transition temperature obtained with theory and experiments are different (i.e., $T_{\rm CDW}^{\rm PBE}=1105$\,K and $T_{\rm CDW}^{\rm exp}=200$\,K). (d) Electronic density of states (DOS), showing the full gap opening as a function of temperature. (e) The phonon dispersion for the $1\times 1$ phase obtained with $T=1600$\,K, $T=1300$\,K, and $T=1150$\,K (from red to blue). (f) The phonon bands for the CDW $2\times 2$ structure obtained at $T=300$\,K.}
\label{fig1}
\end{center}
\end{figure}

The CDW transition in TiSe$_2$ occurs at $T^{\rm exp}_{\rm CDW}\sim 200$\,K\,\cite{disalvo76,chen2015charge}, where the standard $1\times 1$ unit cell is modified into commensurate $2\times 2$ structure with periodic lattice distortions below $T_{\rm CDW}$\,(see Figs.\,S1 and S2 in SI). As shown in Figs.\,\ref{fig1}(a) and \ref{fig1}(b), this structural transition is accompanied by the strong modifications of the electronic band structure near the Fermi level. Namely, the Ti-$3d$ electron-like states that appear at the M point of the Brillouin zone (BZ) of the $1\times 1$ cell, are folded back to the $\Gamma$ point in the CDW $2\times 2$ structure, where they interact strongly with the Se-$4p$ hole-like states. This interaction leads to the opening of the CDW gap between the two Se-$4p$ valence states (denoted v$_1$) and the two out of three Ti-$3d$ conduction bands (denoted c$_2$), where the third one remains almost intact. The temperature dependence of this CDW gap and the total density of states (DOS) are depicted in Figs.\,\ref{fig1}(c) and \ref{fig1}(d), where the relative difference of the CDW gap behaves in accordance to the mean-field theory result for the second-order transition, i.e., as $\Delta E_{c_2-v_1}\propto \tanh({a\sqrt{T_{\rm CDW}/T-1}})$\,\cite{chen2015charge}. 
Since the present results are obtained with the PBE exchange-correlation DFT functional at the harmonic level (i.e., by varying only the electronic entropy) the transition temperature is overestimated $T_{\rm CDW}^{\rm PBE}\approx 1100$\,K. It was demonstrated that a more accurate result could be obtained with a proper inclusion of electron correlation effects\,\cite{hellgren2017critical,novko22}, while to reach the value of $T^{\rm exp}_{\rm CDW}$ it is crucial to incorporate the phonon-phonon corrections\,\cite{zhou20}.
Despite that, the closing of the gap and the relative difference between the low-temperature $T\ll T_{\rm CDW}$ and high-temperature $T> T_{\rm CDW}$ values are in a good agreement with the experimental results as obtained with angle-resolved photoemission spectroscopy\,\cite{monney10} and resonant inelastic x-ray scattering (RIXS) \cite{monney12}. Further, due to the gap opening and modifications of the electronic bands, the DOS is significantly increased at certain energies (e.g., at the gap edges) in accordance to the tunneling spectroscopy studies\,\cite{kolekar18}, which, as we will show below, will lead to the formation of the well-defined (interband) and temperature-dependent absorption peaks in the optical conductivity [see the green and purple arrows in Figs.\,\ref{fig1}(b) and \ref{fig1}(d)].

Several theoretical studies pointed out the importance of using DFT functionals and approaches beyond the semi-local approximations like PBE in order to capture the right electronic structure features of both $1\times 1$ and CDW phases of TiSe$_2$\,\cite{hellgren2017critical,hellgren21,chen23}. However, the CDW gap opening, important for understanding the optical properties of TiSe$_2$, was not thoroughly discussed and compared with the experiments. Our calculations in fact show that the CDW gap opening $\Delta E_{c_2-v_1}$ and its temperature dependence as obtained with the hybrid HSE functional\,\cite{hse} are largely overestimating the experimental results and it is justified to use the semi-local PBE functional to study the optical properties of TiSe$_2$ (see Sec.\,S3 in SI).

The Se-$4p$ states at $\Gamma$ and Ti-$3d$ states around M point are strongly coupled with the acoustic phonon mode at $\mathbf{q}=\mathrm{M}$, which in turn results into significant softening of the latter mode as a function of temperature [see Fig.\,\ref{fig1}(e)]. In our calculations, the soft CDW phonon becomes unstable below $T_{\rm CDW}^{\rm PBE}\approx 1100$\,K, driving the system into the new stable $2\times 2$ configuration with distorted lattice. The CDW $2\times 2$ phase has no soft unstable phonons at low temperatures, as shown in Fig.\,\ref{fig1}(f). 
It should be noted here that the value of the M phonon frequency as obtained with the PBE functional and harmonic DFPT and its functional dependence with respect to $T/T_{\rm CDW}$ are in a good agreement with the experiments\,\cite{hellgren2017critical,novko22}.
As a consequence of this temperature-dependent electronic structure (e.g., opening of the gap) and phonon dynamics (e.g., sensitive soft phonon), TiSe$_2$ is characterized with highly tunable EPC and optical properties.

\begin{figure}[!t]
\begin{center}
\includegraphics[width=\columnwidth]{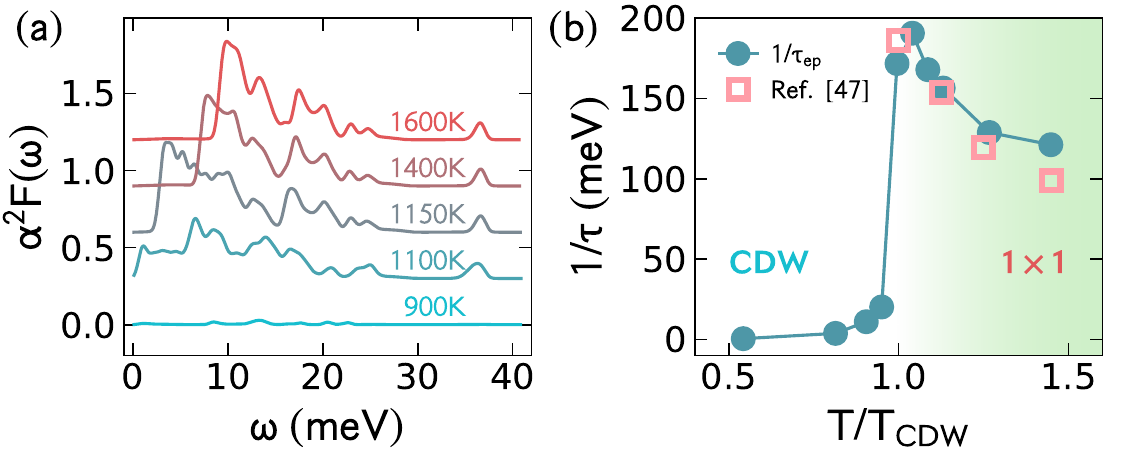}
\caption{(a) Eliashberg electron-phonon spectral function $\alpha^2 F(\omega)$ as a function of the phonon energy and calculated for different temperatures across the CDW transition. The first two light blue lines are for the CDW $2\times 2$ phase, while the rest is calculated for the normal $1\times 1$ structure. (b) The corresponding electron-hole (or optical) scattering rate due to EPC $1/\tau_{\rm ep}$ as a function of temperature (blue dots). $1/\tau_{\rm ep}$ is calculated here in the high-energy regime. Red squares show the results for $1/\tau$ as extracted from the optical conductivity measurements at $T \gtrsim T_{\rm CDW}$\,\cite{velebit16}. Note again the different transition temperature $T_{\rm CDW}$ obtained with theory and experiment (i.e., $T_{\rm CDW}^{\rm PBE}=1105$\,K and $T_{\rm CDW}^{\rm exp}=200$\,K).
}
\label{fig2}
\end{center}
\end{figure}

\begin{figure*}[!t]
\begin{center}
\includegraphics[width=\textwidth]{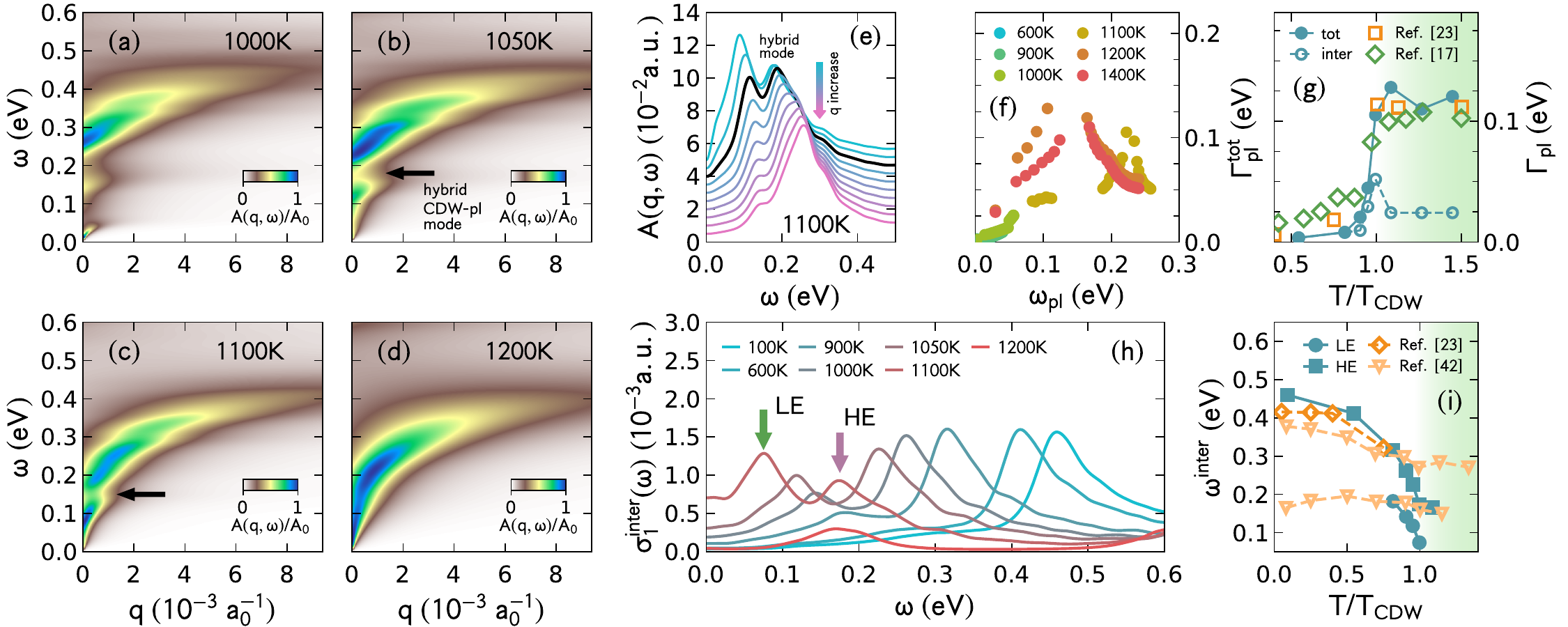}
\caption{Low-energy electron excitation spectra $A(q,\omega)$ of single-layer TiSe$_2$ for several temperatures around $T_{\rm CDW}$: (a) $T=1000$\,K, (b) $T=1050$\,K, (c) $T=1100$\,K, and (d) $T=1200$\,K. The hybrid CDW-plasmon mode is indicated with the black arrows. (e) Spectral function $A(q,\omega)$ at $T=1100$\,K for several momenta $q$ around the hybrid CDW-plasmon mode. (f) Total plasmon damping $\Gamma_{\rm pl}^{\rm tot}$ [i.e., FWHM of plasmon peaks in $A(q,\omega)$], which consists of both interband Landau damping and plasmon-phonon contributions, as a function of plasmon energy $\omega_{\rm pl}$ and temperature across the CDW transition. (g) Total and interband parts of plasmon damping $\Gamma_{\rm pl}$ for fixed plasmon energy $\omega_{\rm pl}$ (i.e., momentum $q$) and as a function of temperature. Plasmon decay rates as obtained from the infrared optical measurments\,\cite{li2007semimetal} and electron energy loss spectroscopy\,\cite{lin22} are shown for comparison.
(h) Real part of the interband optical conductivity $\sigma_1^{\rm inter}(\omega)$ calculated for several temperatures approaching the $T_{\rm CDW}$. The low- (LE) and high-energy (HE) peaks are depicted with green and purple arrows. (i) The energy position of the LE and HE peaks as a function of temperature. The results are compared with infrared spectroscopy\,\cite{li2007semimetal} and RIXS\,\cite{monney12}.
}
\label{fig3}
\end{center}
\end{figure*}

In Fig.\,\ref{fig2}(a) we show the results of the electron-phonon (Eliashberg) spectral function $\alpha^2 F(\omega)$ for the representative temperatures across the CDW transition. The latter spectral function quantifies the degree of the EPC strength for certain phonon energy $\omega$. The overall coupling strength turns out to be quite small for the $2\times 2$ CDW phase below $T_{\rm CDW}$, mostly due to the gap opening and the concomitant small DOS at the Fermi level [see Fig.\,\ref{fig1}(d)]. When the gap is partially closed for $T\lesssim T_{\rm CDW}$ or fully closed for $T>T_{\rm CDW}$, $\alpha^2 F(\omega)$ is significantly increased over the whole phonon-energy range. Particularly strong contribution to the Eliashberg function $\alpha^2 F(\omega)$ (i.e., the low-energy peak) comes from the soft CDW phonon mode around $\mathbf{q}=\mathrm{M}$ point of the Brillouin zone\,\cite{holt01,knowles20}. Note how the phonon dispersion modifications of the soft phonon in Fig.\,\ref{fig1}(e) are reflected in $\alpha^2 F(\omega)$ as a shift of the main peak towards the higher energies. Figure \ref{fig2}(b) shows the corresponding results for the electron-hole (optical) scattering rate due to EPC $1/\tau_{\rm ep}$\,\cite{marsiglio08} calculated in the high-energy limit (i.e., for $\omega$ larger than the highest phonon energy). See SI and Fig.\,S3 for further details. This decay rate can be obtained from $\alpha^2 F(\omega)$ and it is an integral part of the Drude optical conductivity\,\cite{kupcic14}, describing the scattering processes between the screened intraband electron-hole excitations (e.g., plasmons) and phonons\,\cite{novko17}. Following the temperature behavior of $\alpha^2 F(\omega)$, the calculated scattering rate $1/\tau_{\rm ep}$ shows a dramatic transition from the low values (below 10\,meV) at $T<T_{\rm CDW}$ to the large-damping region for $T\gtrsim T_{\rm CDW}$. This result agrees well with the optical scattering rate obtained in the optical conductivity measurements for the high-temperature phase $T\gtrsim T_{\rm CDW}$\,\cite{velebit16}. In order to explain this intense and highly temperature-dependent damping rate, the authors of Ref.\,\citenum{velebit16} discussed the results in terms of strong inter-valley scatterings (between the Se-$4p$ hole and Ti-$3d$ electron pockets) mediated by the soft CDW-related phonon mode. Our calculations support this picture. 
Note once again that the current results for $1/\tau_{\rm ep}$ are obtained with harmonic linear EPC calculations and we do not expect that the corresponding functional dependence with respect to $T/T_{\rm CDW}$ will be modified significantly with the inclusion of anharmonic corrections. Except approaching the experimental value of $T_{\rm CDW}$, we only expect that the lattice fluctuations will slightly reduce the current value of $1/\tau_{\rm ep}$\,\cite{hui74}.


The calculated optical excitation properties of single-layer TiSe$_2$ are presented in Fig.\,\ref{fig3}. The low-energy electron excitation spectra $A(q,\omega)$ as a function of momentum $q$ and for various temperatures near the $T_{\rm CDW}$ are depicted in panels (a)-(d). The spectra are dominated by the 2D plasmon mode characterized with a $\sqrt{q}$ dispersion and screened interband excitations. 
As the temperature approaches $T_{\rm CDW}$ the energy of the 2D plasmon increases (for all momenta $q$). Similar observation was reported for bulk TiSe$_2$, where a bulk three-dimensional plasmon with a gap at $q=0$ is considerably blueshifted with temperature around $T_{\rm CDW}$\,\cite{li2007semimetal,lin22}. This effect comes from the abrupt increase of the electronic DOS at the Fermi level and consequently the increase of the Drude weight and effective number of carriers\,\cite{knowles20} once the gap is closed [see Fig.\,\ref{fig1}(d)]. For $T<T_{\rm CDW}$ the 2D plasmon and CDW interband excitations are distinguished and energetically well separated, while slightly below the $T_{\rm CDW}$ the plasmon interacts with the CDW excitations forming a mixed hybrid mode (see also Fig.\,S4 in the SI for the excitation spectra across the larger temperature range). For $T=1050$\,K and $T=1100$\,K this coupling is manifested as an avoided crossing between the two modes [see the black arrows in Figs\,\ref{fig3}(b) and \ref{fig3}(c)]. The spectral lineshape of this hybrid CDW-plasmon mode is demonstrated further in Fig.\,\ref{fig3}(e), where we show $A(q,\omega)$ near the hybridization energy for $T=1100$\,K. The strong deviation from the non-interacting Lorentzian lineshape and the appearance of the two-peak structure is evident. A strikingly similar hybrid CDW-plasmon mode was recently observed in $2H$-TaSe$_2$ van der Waals thin films by means of Fourier-transform infrared spectroscopy\,\cite{song2021plasmons}. This agreement shows that the hybrid CDW-plasmon modes might be universal in the CDW-bearing quasi-2D materials, since the 2D plasmon modes are usually well-defined in the low-energy region where the CDW excitations are possible. On the other hand, bulk plasmons are less dispersive and have finite energy at the long wavelengths, and therefore the direct CDW-plasmon coupling is less probable in the bulk CDW materials\,\cite{lin22}.

We are now in the position to inspect the total contribution to the plasmon damping $\Gamma_{\rm pl}$, which consists of both interband Landau damping and plasmon-phonon parts, and analyze its temperature dependence. 
The total damping is obtained from $A(q,\omega)$ by fitting Lorentzians to the plasmon peaks and extracting the corresponding full-width-at-half-maximum (FWHM).
From Fig.\,\ref{fig3}(f) we see that the total plasmon damping overall increases with increasing energy (and hence momentum $q$), which was also observed for the 2D plasmon in TaSe$_2$\,\cite{song2021plasmons}, and partially comes from the 2D nature of Coulomb screening for which $\Gamma_{\rm pl}\propto q$\,\cite{kupcic2014damp}. For temperatures where the plasmon is well formed and ranges up to 0.3-0.4\,eV (i.e., $T>1100$\,K), the plasmon damping $\Gamma_{\rm pl}$ shows intense peaks that come from the Landau damping due to the CDW interband excitations.

Interestingly, for the selected plasmon energy $\omega_{\rm pl}$ (i.e., momentum $q$) $\Gamma_{\rm pl}$ shows an unconventional temperature dependence, i.e., it rapidly increases around $T_{\rm CDW}$ [Fig.\,\ref{fig3}(g)]. We explain this anomalous dependence as a consequence of the interplay between the strong plasmon-phonon scattering and Landau damping due to CDW gap excitations. From Fig.\,\ref{fig2}(b) we see that the scattering rate due to EPC abruptly increases around $T_{\rm CDW}$, which then contributes to the plasmon decay. As one can see from Fig.\,\ref{fig3}(g) (open blue symbols) a purely interband contribution to $\Gamma_{\rm pl}$ is strongest around the $T_{\rm CDW}$. However, note that the EPC contribution to $\Gamma_{\rm pl}$ is much larger than the Landau damping coming from the CDW gap excitations. Considering the general importance of the EPC in TiSe$_2$, e.g., for the formation of CDW and superconductivity\,\cite{yoshida80,calandra11}, it is not entirely surprising that it plays an important role in plasmon dynamics. We point out that the similar mechanism might also be behind the anomalous temperature dependence of 2D plasmon damping in the few-layer TaSe$_2$\,\cite{song2021plasmons}.

The presented result for plasmon damping $\Gamma_{\rm pl}$ agrees fairly well with the decay rate for the bulk plasmon as obtained with the infrared optical measurements\,\cite{li2007semimetal} and electron energy loss spectroscopy\,\cite{lin22}. In Ref.\,\citenum{li2007semimetal} a drastic decrease of $\Gamma_{\rm pl}$ below $T_{\rm CDW}$ was explained in terms of a reduced scattering phase space due to opening of the CDW gap, while no discussion was provided on the relevant scattering channel. The same effect was elaborated more in Ref.\,\citenum{lin22}, where the strong modifications of $\Gamma_{\rm pl}$ was attributed to suppression (enhancement) of the Landau damping due to interband excitations over the CDW gap below (above) $T_{\rm CDW}$. Note however that they used a phenomenological model to introduce the CDW interband excitations into the total response function, and electron-phonon scattering channel was not taken into account. The results presented in Fig.\,\ref{fig3} suggest, on the other hand, that the alterations of $\Gamma_{\rm pl}$ dominantly come from the EPC, in particular, from the inter-valley electron-hole scatterings assisted by the soft CDW-related acoustic phonon at $\mathbf{q}=\mathrm{M}$. It is important to note here that the bulk plasmon energy in TiSe$_2$ ranges from 50\,meV for $T<T_{\rm CDW}$ to 150\,meV for $T \gtrsim T_{\rm CDW}$\,\cite{li2007semimetal,kogar17,lin22}, which is close to the energy range obtained here for the 2D plasmon mode of TiSe$_2$ monolayer. This suggests that the CDW gap excitations are unlikely to explain the sudden increase of the plasmon linewidth above $T_{\rm CDW}$ and that the EPC is probably a dominant plasmon decay channel also in the bulk TiSe$_2$, as already indicated in Fig.\,\ref{fig3}(g).

To comprehend the CDW-plasmon coupling more closely, we also calculate the real part of the interband optical conductivity $\sigma_1^{\rm inter}(\omega)$ for several relevant temperatures [Fig.\,\ref{fig3}(h)]. The results around the $T_{\rm CDW}$ are characterized with the two well-defined low- (LE) and high-energy (HE) interband excitation peaks, which correspond to the CDW gap excitations pointed out in Fig.\,\ref{fig1} with the green and purple arrows. As shown in Fig.\,\ref{fig3}(i), the energy position of these interband peaks are in a fair agreement with the infrared spectroscopy\,\cite{li2007semimetal} and RIXS\,\cite{monney12} measurements. Ref.\,\citenum{li2007semimetal} reported only one peak that decreases with temperature towards $T_{\rm CDW}$ and corresponds to the denoted HE peak (see Figs.\,S5 and S6 in SI for further comparison). On the other hand, the RIXS measurements\,\cite{monney12} were able to detect two excitations related to the CDW gap opening, which we attribute to the LE and HE transitions. Note that while the HE peak from Ref.\,\citenum{monney12} agrees well with the Ref.\,\citenum{li2007semimetal} and our results, the LE peak from RIXS does not exactly follow the present results. The latter could be partially understood by considering that the expression for simulating the RIXS is slightly different than the optical conductivity formula as well as that the LE peak in RIXS is close to the elastic line and therefore difficult to discern\,\cite{monney12}. It is also important to emphasize here that the energy position of the HE peak is largely overestimated when the hybrid HSE functional is used (see Fig.\,S7 in SI).
As the temperature increases the two peaks shift to the lower energies and are finally suppressed around and above $T_{\rm CDW}$, since the CDW gap is closed. Further, from the results of $\sigma_1^{\rm inter}(\omega)$ it is clear that depending on the temperature, the 2D plasmon mode interacts with one of these CDW electron excitations, which in turn brings about the two pronounced optical features, i.e., hybrid CDW-plasmon mode and interband Landau damping.

In summary, we have utilized a well-established \emph{ab-initio} electromagnetic linear response formalism based on DFT and DFPT in order to investigate the dynamics of 2D plasmon mode across the CDW transition in single-layer TiSe$_2$. We have disentangled various scattering mechanisms, like CDW gap excitation and plasmon-phonon coupling, and have uncovered an intriguing unconventional temperature dependence of the plasmon broadening in 2D TiSe$_2$. Below $T_{\rm CDW}$, we have found a highly tunable hybrid mode that comes from the coupling between the CDW gap excitations and 2D plasmon.
The present study demonstrates a great potential of the correlated TMDs as a highly tunable plasmonic materials with unconventional optical features, induced by the charge-ordered states. Having in mind the rich phase diagram of TiSe$_2$ it is expected that the CDW-plasmon coupling could be further controlled with carrier doping\,\cite{morosan2006superconductivity,luo16} and pressure\,\cite{kusmartseva2009pressure}, or even additional hybrid modes might be activated as a consequence of the interactions with superconductivity-related excitations\,\cite{costa21}. Note that similar features are expected in other 2D semi-metallic and metallic TMDs hosting a CDW order, such as ZrTe$_2$\,\cite{song23}, TiTe$_2$\,\cite{antonelli22}, or NbSe$_2$\,\cite{ugeda15}. Finally, our work supports the idea that plasmon modes could be utilized as a loupe for tracking ordered phases and the corresponding intrinsic excitation mechanisms in correlated systems\,\cite{song2021plasmons,costa21,vanloon14,asmara17,gao22}.

\begin{acknowledgement}
D.N. acknowledges financial support from the Croatian Science Foundation (Grant no. UIP-2019-04-6869) and from the European Regional Development Fund for the ``Center of Excellence for Advanced Materials and Sensing Devices'' (Grant No. KK.01.1.1.01.0001). Z.T. acknowledges financial support from the Iran Science Elites Federation. Part of the computational resources were provided by the DIPC computing center.
\end{acknowledgement}

\begin{suppinfo}
More information on theoretical methods, computational details, crystal structure of the $2\times2$ phase, periodic lattice distortions, crucial steps for obtaining the EPC contribution to the electron-hole scattering rate, comparison between the calculated optical conductivity and experiment, as well as impact of the non-local hybrid functionals and spin-orbit coupling on the optical properties of TiSe$_2$.\\
\end{suppinfo}

\bibliography{tise2}

\end{document}